\documentclass[aps,prc,twocolumn,showpacs,amsmath,amssymb,preprintnumbers]{revtex4}

\pdfoutput=1   
\usepackage[10pt]{type1ec}  % use only 10pt fonts
\usepackage[T1]{fontenc}
\usepackage{CJKutf8}
\newenvironment{SChinese}{%
  \CJKfamily{gbsn}%
  \CJKtilde
  \CJKnospace}{}
\usepackage{latexsym} % Gets \Box etc
\usepackage{graphicx}       % For PostScript figures
\usepackage{color}
\usepackage{bm}   % Bold math: \bm{1}

\newcommand{\al}{\alpha}

\newcommand{\de}{\delta}
\newcommand{\De}{\Delta}
\newcommand{\ep}{\varepsilon}
\newcommand{\eps}{\epsilon}

\newcommand{\la}{\lambda}

\newcommand{\si}{\sigma}

\newcommand{\beq}{\begin{equation}}
\newcommand{\eeq}{\end{equation}}
\newcommand{\ba}{\begin{array}}
\newcommand{\ea}{\end{array}}
\newcommand{\bea}{\begin{eqnarray}}
\newcommand{\eea}{\end{eqnarray}}
\newcommand{\bi}{\begin{itemize}}  %\setlength{\itemsep}{0\parsep}}
\newcommand{\ei}{\end{itemize}}
\newcommand{\ben}{\begin{enumerate}} %\setlength{\itemsep}{0\parsep}}
\newcommand{\een}{\end{enumerate}}
\newcommand{\bc}{\begin{center}}
\newcommand{\ec}{\end{center}}

\newcommand{\p}{\partial}

\newcommand{\txt}{\textstyle}
\newcommand{\dsp}{\displaystyle}
\newcommand\eqn[1]{(\ref{#1})}      % parentheses around the LaTex "ref" macro
\newcommand{\half} {{\txt \frac{1}{2}}}

\newcommand{\m}{{\rm m}}
\newcommand{\fm}{{\rm fm}}

\newcommand{\MeV}{{\rm MeV}}

\newcommand{\chiQ}{\raisebox{0.2ex}{$\chi$}^{\phantom{y}}_Q}
\newcommand{\nQ}{n^{\phantom{y}}_{\!Q}} % phantom y pushes Q subscript down
\newcommand{\lambdaD}{\lambda_D}

\newcommand{\mucrit}{\mu_{\rm crit}}
\newcommand{\sicrit}{\si_{\rm crit}}

\newcommand{\Rcell}{{R_{\rm cell}}}

\newcommand{\pcell}{p_{\rm cell}}
\newcommand{\pQM}{p^{\phantom{y}}_{\rm QM}} % phantom y pushes subscript down
\newcommand{\mue}{\mu_{e}}

\begin{document}

\title{Strangelet dwarfs}
\author{Mark G. Alford, Sophia Han
(\begin{CJK}{UTF8}{}\begin{SChinese}韩 君\end{SChinese}\end{CJK})
}

\affiliation{Physics Department, Washington University,
St.~Louis, MO~63130, USA}
\author{Sanjay Reddy}
\affiliation{
Institute for Nuclear Theory, University of Washington, 
Seattle, Washington 98195-1550, USA
}

\begin{abstract}
If the surface tension 
of quark matter is low enough, quark matter is not self bound.
At sufficiently low
pressure and temperature, it will take the form of
a crystal of positively charged strangelets in a 
neutralizing background of electrons.
In this case there will exist, 
in addition to the usual family of strange stars,
a family of low-mass large-radius objects
analogous to white dwarfs, which we call ``strangelet dwarfs''.
Using a generic parametrization of the equation of
state of quark matter, we calculate the mass-radius
relationship of these objects.
\end{abstract}

\date{7 March 2012}

\preprint{INT-PUB-11-054}


\maketitle

\section{Introduction}
\label{sec:intro}

The matter that is directly observed in nature
consists of atoms, whose nuclei are droplets of nuclear matter
composed of up and down quarks. Nuclear matter is very stable:
the most stable nuclei have lifetimes longer than the age of the universe.
However, it has been hypothesized 
\cite{Bodmer:1971we,Witten:1984rs,Farhi:1984qu}
that nuclear matter may actually be metastable, and the
true ground state of matter consists of a combination of roughly
equal numbers of up, down, and strange quarks known as ``strange matter''.
Strange matter is hypothesized to exist as
(kilometer-sized) pieces, known as ``strange stars'' (reviewed in
\cite{Weber:2004kj}), or as small nuggets, known as ``strangelets''
\cite{Farhi:1984qu}.  
It has further been hypothesized that dark matter could be some 
form of quark matter, trapped in strangelets or strange stars before
the era of nucleosynthesis
\cite{Zhitnitsky:2002qa,Banerjee:2002et,Forbes:2009wg}. 
However, even if strange matter is not invoked as a dark matter candidate,
there could still be a population of strange matter objects, from strange
stars to strangelets, many of which
would be relatively non-luminous. In this article we
show that the masses and radii of such objects can extend in to the range
expected for planets.  Recent surveys such as the Microlensing Observations in Astrophysics (MOA) and 
the Optical Gravitational Lensing Experiment (OGLE) to detect such low mass non-luminious low mass objects 
by gravitational lensing have yielded interesting results \cite{Sumi:2011}. 
The hypothetical compact objects we predict 
could be detected by such methods and such surveys could place stringent bounds or perhaps hint at their 
possible existence.

It is generally assumed that strange stars are compact objects, with sizes in
the 10 kilometer range \cite{Weber:2004kj}, ending at a sharp surface of
thickness $\sim$ 1 Fermi, perhaps with a very thin electrostatically suspended
nuclear matter crust~\cite{Alcock:1986hz,Stejner:2005mw,Usov:1997eg}.
However, if the surface tension $\si$ of the interface between quark matter
and the vacuum is less than a critical value $\sicrit$ 
(of order a few MeV/fm$^2$ in typical models of quark matter)
then large strangelets
are unstable against fission into smaller ones
\cite{Jaikumar:2005ne,Alford:2006bx,Alford:2008ge}, and the energetically
preferred state is a crystal of strangelets: a mixed phase consisting of
nuggets of positively-charged strange matter in a neutralizing background of
electrons.

In this ``low surface tension'' scenario, strange stars are {\em not}
self-bound: they require gravitational attraction to bind the strangelets.
Stars made of strange matter are then qualitatively
similar to those made of nuclear matter:
in each case the mass-radius relation has two branches, one compact
and the other diffuse.
For nuclear matter, the compact branch contains neutron stars, which
consist of gravitationally bound nuclear matter, with an outer
crust that is a crystal of nuclei in a background of
electrons; the diffuse branch contains white
dwarfs, which are a gravitationally bound cold plasma of 
nuclei (ions) and electrons, forming,  at sufficiently low temperature,
a crystalline structure
(see, e.g., \cite{Potekhin:2000}).
For strange matter with a low surface tension, there are similarly two
branches. The compact
branch contains strange stars with a crust that consists of
strangelets in a background of electrons; this ``strangelet crystal crust''
was studied in Ref.~\cite{Alford:2008ge}. In this paper we
study the diffuse branch, which has no core of uniform quark matter: these
stars consist entirely of strangelets in a background of degenerate electrons,
so by analogy with white dwarfs we call them strangelet dwarfs.

The strangelet-crystal phase is a charge-separated phase.
Charge separation is favored by the internal energy
of the phases involved, because a neutral phase is always
at a maximum of the free energy with respect to the electrostatic
potential (see \cite{Ravenhall:1983uh,Glendenning:1992vb};
for a pedagogical discussion see \cite{Alford:2004hz}).
The domain structure is determined by competition between
surface tension (which favors large domains) and electric field
energy (which favors small domains). Debye screening plays
a role in determining the domain structure, 
because it redistributes the electric charge, concentrating it
in the outer part of the quark matter domains and the inner part
of the surrounding electron gas, and thereby modifying the internal energy
and electrostatic energy contributions.
Our parameterization \eqn{eqn:EoS}
of the electrostatic properties of quark matter
 is generic, but is not appropriate
for strangelets in the color-flavor locked (CFL) phase \cite{Alford:1998mk}, 
which is a degenerate
case requiring separate treatment (see Sec.~\ref{sec:params}).

To obtain the $M(R)$ relation of strangelet dwarfs,
we solve the Tolman Oppenheimer Volkoff
equation \cite{Tolman:1939jz,Oppenheimer:1939ne}, using 
the equation of state of the mixed phase. We obtain the equation of state
by assuming that the strangelet lattice can be divided into unit cells
(``Wigner-Seitz cells'') and calculating the pressure of a
cell as a function of its energy density.  
Our approach is similar to that used in previous studies of
the strangelet crystal \cite{Alford:2006bx,Alford:2008ge} (except that
in this paper we include electron mass effects)
and in studies of mixed phases
of quark matter and nuclear matter in the interior of neutron stars
\cite{Maruyama:2007ey}.

The main assumptions that we make are:\\
1) We assume that the strangelets in the plasma form a regular lattice
of Wigner-Seitz cells, which we treat as rotationally invariant
(spherical). In reality the cells will be unit cells of some
regular lattice. We do not consider lower-dimensional structures
(rods or slabs) because in Ref.~\cite{Alford:2008ge} we found that such
structures were never energetically favored.\\
2) Within each Wigner-Seitz cell we use a Thomas-Fermi approach,
solving the Poisson equation to obtain the charge distribution,
energy density, and pressure. This is incorrect for
very small strangelets, where the energy level structure
of the quarks becomes important \cite{Madsen:1994vp,Amore:2001uf}.
\\
3) We treat the interface between quark matter and the vacuum as a
sharp interface which is characterized
by a surface tension. We assume there is no charge localized on 
the surface.
(Thus we neglect any surface charge that might arise
from the reduction of the density
of states of strange quarks at the surface 
\cite{Madsen:2000kb,Madsen:2001fu,Madsen:2008bx,Oertel:2008wr}.)
\\
4) We neglect the curvature energy of a
quark matter surface \cite{Christiansen:1997rc,Christiansen:1997vt},
so we do not allow for ``Swiss-cheese'' mixed phases, in which 
the outer part of the Wigner-Seitz cell is filled with quark matter, with
a cavity in the center, for which the curvature
energy is crucial.
\\
5) We work at zero temperature. 

In our calculations we use units $\hbar=c=\eps_0=1$, 
so $\al=e^2/(4\pi)\approx 1/137$. 

%The temperature of the surface of a compact star, even during a flare
%\cite{Lyubarsky:2002cs}, is expected to be less than 100 keV, so we expect
%this to be a reasonable approximation.

\section{Phenomenological description of quark matter}
\label{sec:characterization}

We use the fact that in most phases of quark matter
the chemical potential for negative electric charge $\mue$ 
is much less than the chemical potential for quark number $\mu$. 
This allows us to write down a model-independent parameterization of
the quark matter equation of state, expanded
in powers of $\mue/\mu$ \cite{Alford:2006bx},
\beq
\pQM(\mu,\mue)  \approx
 p_0(\mu)-\nQ(\mu)\mue + \half\chiQ(\mu) \mue^2+\ldots
\label{eqn:EoS}
\eeq
Note that the contribution of electrons to the pressure 
of quark matter is ${\cal O}(\mue^4)$, and is neglected.
This is a very good approximation for small 
strange quark mass, which corresponds to
small $\nQ$. 
(For the largest value of $\nQ$
that we study, $\mue$ in neutral quark matter 
is close to 100 MeV, and the assumption is still reasonable.)

As noted in Sec.~\ref{sec:intro}, we assume that
the interface between quark matter and vacuum has a
surface tension $\si$, and we neglect any curvature energy.

The quark density $n$ and the electric charge density 
$q^{\phantom{y}}_{\rm QM}$ 
(in units of the positron charge) are
\beq
%\ba{rcl}
n = \frac{\p \pQM}{\p \mu},\qquad
q^{\phantom{y}}_{\rm QM} =  -\frac{\p \pQM}{\p \mue}
 = \nQ - \chiQ\mue \ .
%\ea
\label{charges}
\eeq
So in uniform neutral quark matter the electron chemical
potential is $\mue^{\rm neutral}=\nQ/\chiQ$. Eq.~\eqn{eqn:EoS}
is a generic parametrization if
$\mue^{\rm neutral}\ll\mu$, which is typically the case in three-flavor
quark matter.

The bag constant enters in $p_0(\mu)$, and we will fix it by
requiring that the first-order 
transition between neutral quark matter and the vacuum
occur at quark chemical potential $\mucrit$, 
i.e.~$p(\mucrit,\mue^{\rm neutral})=0$.
Because we
are assuming that the strange matter hypothesis is valid, we require
$\mucrit \lesssim 310$~MeV, since at $\mu\approx 310$~MeV there is
a transition from vacuum to neutral nuclear matter.
In this article we will typically use $\mucrit=300~\MeV$. 
The value of $\mu$ inside our quark matter lumps will always be very close
to $\mucrit$, so we can also expand in powers of $\mu-\mucrit$, and write
%evaluate $\nQ$ and $\chiQ$ at $\mucrit$, and not
%be concerned about their $\mu$-dependence.
%so we can treat the charge density parameter $\nQ$ and the 
%charge susceptibility $\chiQ$ as constants, and we can write
\beq
\ba{rcl}
\pQM(\mu,\mue)  &\approx&\dsp
n\,(\mu-\mucrit) + \half\chi\,(\mu-\mucrit)^2 \\ 
&& + \frac{n_Q^2}{2\chiQ} -\nQ\mue + \half \chiQ\mue^2 \ .
\ea
\label{generic_EoS}
\eeq
A quark matter equation of state can then be expressed in terms of
6 numbers: $\mucrit$,
the charge density $\nQ$ and charge susceptibility $\chiQ$
evaluated at $\mu=\mucrit$, 
the quark number density $n$ and susceptibility $\chi$
evaluated at $\mu=\mucrit$, and the surface tension $\si$.

We will restrict ourselves
to values of the surface tension that are below the critical value
\cite{Alford:2006bx} 
\beq
\ba{rcl}
\sicrit &=&\dsp 0.1325 \,
\frac{n_Q^2 \lambda_D}{\chi_Q}
=0.1325\,\frac{n_Q^2}{\sqrt{4\pi\alpha}\chi_Q^{3/2}},
\ea
\label{sicrit_result}
\eeq
where $\lambda_D$ is the Debye screening length in quark matter
\beq
\la_D = \frac{1}{\sqrt{4\pi\alpha\chiQ}} \ .
\label{Debye_length} 
\eeq 
If the surface tension is larger than $\sicrit$
then the energetically favored structure at low pressure
will not be a strangelet crystal, and there will be
no strangelet dwarfs.
% \cite{Alcock:1986hz,Usov:1997eg}.
Rough estimates of surface tension from the
bag model are in the range $4$ to $10~\MeV/\fm^2$ 
\cite{Berger:1986ps,PhysRevC.44.566.2}, and
for typical models of quark matter,
$\sicrit$ is of order $1$ to 
$10~\MeV/\fm^2$ \cite{Alford:2006bx}, so it is reasonable to
explore the possibility that strange quark matter could have a surface tension
below $\sicrit$.

\subsection{Specific equations of state}
\label{sec:specific}

When we show numerical results we will need to vary $\nQ$ and $\chiQ$
over a range of physically reasonable values. To give a rough idea
of what values are appropriate, we consider the example of
non-interacting three-flavor quark matter, for which $\nQ$ and $\chiQ$
become functions of $\mu$ and the strange quark mass
$m_s$, while $p_0$ is in addition
a function of the bag constant $B$. Expanding to lowest non-trivial
order in $m_s$,
\beq
\ba{rcl}
p_0(\mu) &=& \dsp \frac{9\mu^4}{12\pi^2} -B  \ ,\\[2ex]
\nQ(\mu,m_s) &=& \dsp \frac{m_s^2\mu}{2\pi^2} \ ,\\[2ex]
\chiQ(\mu,m_s) &=& \dsp \frac{2\mu^2}{\pi^2}.
\label{unpaired}
\ea
\eeq
We emphasize that these expressions are simply meant to give a rough
idea of reasonable physical values for $\nQ$ and $\chiQ$. Our treatment
does not depend on an expansion in powers of $m_s$.
To tune the transition between neutral quark matter and the vacuum
so it occurs at $\mu=\mucrit$ (see previous subsection), we set $B$ so that
$p_0(\mucrit)=\half n_Q^2(\mucrit)/\chiQ(\mucrit)$.

In the regions between lumps of strange matter, we will assume that
there is a degenerate  electron gas, whose pressure, and charge
density in units of $e$, are
\beq
\ba{rcl}
p_{e^{\!-}}(\mue) &=& \dsp \frac{1}{24\pi^2}\biggl( 
   ( 2 k_{Fe}^2 - 3m^2) k_{Fe} \mue  \\
  &&\dsp +\ 3 m^4 \ln\Bigl(\frac{k_{Fe}+\mue}{m}\Bigr) \biggr) \ ,\\
q_{e^{\!-}}(\mue) &=& \dsp -\frac{1}{3\pi^2}k_{Fe}^3 \ .
\ea
\label{electrons}
\eeq
where $\mue^2 = k_{Fe}^2 + m_e^2$.
Note that at low pressures this is
more accurate than the electron gas equation of state
used in Ref.~\cite{Alford:2008ge}, where the electron mass was set to zero.

\section{Equation of state of strangelet crystal}
\label{wigner-seitz}

\subsection{Wigner-Seitz cell}
Following the approach of \cite{Alford:2008ge}, we analyze a 
spherical Wigner-Seitz cell of radius $\Rcell$, with a
sphere of quark matter at the center of radius $R$. 
We use the Thomas-Fermi approximation to calculate $\mue(r)$,
\beq
% \nabla^2 \phi(r) &=& - \rho(r) \ , \\
\nabla^2 \mue(r) = -4 \pi \alpha q(r) \ ,
\label{Poisson}
\eeq
where $q(r)$ is the electric charge density in units of the positron charge $e$,
and $\mue$ is the electrostatic potential divided by $e$.

The boundary conditions are that there is no electric field in the
center of the cell (no $\de$-function charge there), and
no electric field at the edge of the cell (the cell is electrically neutral),
\beq
\frac{d \mue}{d r}(0) = 0 \ , \qquad
\frac{d \mue}{d r}(R_{\rm cell}) = 0 \ . 
\label{BC}
\eeq
We also need a matching condition at the edge of the quark matter.
Since we assume that no charge is localized on the surface, we
require continuity of $\mue$ and its first derivative (the electric field)
at $r=R$.

The value of $\mu$ inside the strange matter will be slightly
different from $\mucrit$ because the surface
tension compresses the droplet.
To determine the value of $\mu$, we require the pressure discontinuity
across the surface of the strangelet to be balanced by the surface tension:
\beq
\pQM(\mu,\mue(R)) -  p_{e^{\!-}}(\mue(R)) = \frac{2\si}{R} \ .
\label{pressure_disc}
\eeq

Once these equations are solved, we can obtain the 
equation of state of matter made of such cells.
The total energy of a cell is
\beq
\ba{rcl}
E &= & \dsp 4\pi \int_0^R \! r^2 dr \, \Bigl(
  \mu n(\mue) - \half \mue q^{\phantom y}_{\rm QM}(\mue) 
% Need phantom y to push subscript down: stupid TeX.
  - \pQM(\mu,\mue)\Bigr)  \\[3ex]
 &+& \dsp 4\pi \int_R^{\Rcell} \!r^2 dr \,
    \Bigl( - \half \mue q_{e^{\!-}}(\mue) - p_{e^{\!-}}(\mue) \Bigr) \\[3ex]
 &+& \dsp 4\pi  R^2\, \si \ ,
\ea
\label{energy}
\eeq
The $-\half \mue q$ terms in \eqn{energy} 
come from combining $-\mue q$ (from the relationship
between energy density and pressure) with the electric field energy density
$+\half \mue q$.
The pressure of the cell is simply the pressure of the electrons
at the edge of the cell,
\beq
\pcell = p_{e^{\!-}}\bigl(\mue(\Rcell)\bigr) \ .
\label{pext}
\eeq
The total number of quarks is
\beq
N = 4\pi \int_0^R \! r^2 dr \, n(\mu,\mue) \ .
\eeq
The volume of the cell is $V= (4/3)\pi\Rcell^3$.

By varying $R$ and $\Rcell$ we generate a two-parameter family of
strangelets. However, there is really only a single-parameter
family of physical configurations, parameterized by the external
pressure $\pcell$. On each line of constant $\pcell$ in the
$(R,\Rcell)$ parameter space, we must minimize the enthalpy per quark,
\beq
h = \frac{E + \pcell V}{N} \ ,
\label{eq:enthalpy}
\eeq
to find the favored value of $R$ and $\Rcell$.
We assume zero temperature so $h$ is also the
Gibbs free energy per quark.

We now have a well-defined way to obtain the equation of state
of the mixed phase of quark matter, namely
the energy density $\ep=E/V$ as a function
of the pressure $\pcell$. 

\subsection{Numerical solution}
\label{sec:exact}

Inside the quark matter, the solution to the Poisson equation \eqn{Poisson}
that obeys the boundary condition at the origin is
\beq
\mu_{e}(r)=\frac{n_Q}{\chi_Q}+
 \frac{A}{r \lambda_D} \sinh(\frac{r}{\lambda_D}) \ ,
\label{mue-qm}
\eeq
where $A$ will be determined by matching conditions.

In the degenerate electron gas region outside the strange matter, 
from \eqn{electrons} and \eqn{Poisson} the
Poisson equation becomes 
\beq
\nabla^2 \mue(r) = \frac{4\alpha}{3\pi} (\mue^2-m_e^2)^{3/2} \ ,
\label{eqn:Poisson-vac}
\eeq
which must be solved numerically. For a given value of $A$ we 
find from \eqn{mue-qm} the value and slope of $\mue(r)$ at $r=R$,
and use these as initial values to propagate $\mue(r)$ out to
$r=\Rcell$ using \eqn{eqn:Poisson-vac}.
We vary $A$ until we obtain a solution that
obeys the boundary condition of no electric
field at the edge of the cell.

\subsection{Low-pressure approximations}
\label{sec:lowp}

If the pressure is not too high, the strangelet crystal consists of large
Wigner-Seitz cells ($\Rcell\gg R$). 
In this regime one can obtain approximate analytic expressions for the equation
of state of the crystal by assuming that the electrons have a roughly
constant density outside the strangelet. We give these expressions below,
and in later sections we use them to calculate mass radius relations
for large strangelet dwarf stars.
However, we expect these approximations break down at ultra-low
pressures, when the cell size becomes so
large that screening cannot be ignored, and the electrons are
clumped around the strangelets, forming atoms, rather then being
roughly uniformly distributed between the strangelets. This will
happen when $\Rcell$ approaches the Bohr radius 
$a_0=1/(\al m_e)$, i.e. when $\pcell \lesssim \al^5 Z^{5/3} m_e^4 
\approx (10^{-12} \MeV^4) Z^{5/3}$. At these ultra-low pressures
one should use an atomic matter equation of state: we do not do this,
since we expect it will only affect a very small surface layer of the star, 
without any appreciable effect on the mass-radius relationship.

The equation of state $\ep(\pcell)$ is found by writing 
the energy density $\ep$ of the cell and its pressure as a function
of the size of the cell. For now we will 
treat the size $R$ and charge $Z$ of the central strangelet as unknowns;
later we will estimate their values.

Since the pressure inside a large cell is very low the energy density
of the quark matter is approximately $n \mucrit$, so
\beq
\ep \approx n \mucrit \frac{R^3}{\Rcell^3} \ .
\label{e-bigcell}
\eeq

To obtain the pressure at the edge of the cell we need to estimate the density
distribution of the electrons outside the strangelet. 

\subsubsection{Constant potential approximation}
\label{sec:constapprox}

The simplest approximation is to ignore screening, taking the
electron Fermi momentum $k_{Fe}$ to be % $\mue$ 
independent of $r$ outside the strangelet 
(Sec.~I of Ref.~\cite{Salpeter:1961zz}). 
Imposing neutrality of the cell fixes
the Fermi momentum of the electrons,
\beq
k_{Fe}^3 = \frac{9\pi Z}{4\Rcell^3} \ .
\label{kFe-const}
\eeq
Using \eqn{e-bigcell}, we obtain the equation of state $\ep(\pcell)$
of the strangelet crystal
\beq
\ep \approx \mucrit n \frac{ 4 (k_{Fe} R)^3}{9\pi Z}
% \frac{n \mucrit k_{Fe}^3}{3 \pi^2 \nQ \xi(R/\lambdaD)} \ ,
\label{eos-const}
\eeq
where  we use \eqn{electrons} to relate the
electron Fermi momentum to $\pcell$.

Because the constant potential approximation gives a fairly simple expression
we can use it to understand how the strangelet crystal EoS
depends on the parameters of the quark matter EoS, and hence how the
$M(R)$ curve for strangelet dwarf stars depends on those parameters.
Note that in \eqn{eos-const} the dependence of the energy density 
on the pressure is via a universal and monotonically
increasing function $k_{Fe}(p)$;
dependence on the quark matter parameters
enters via the factor that multiplies this function. To make the
dependence on quark matter parameters explicit we
use results for $R$ and $Z$ from Sec.~\ref{sec:RZ} below,
and rewrite \eqn{eos-const} for the EoS of the strangelet crystal as
\beq
\ba{rcl}
\ep(\pcell) &\sim& S\,  \bigl(k_{Fe}(\pcell)\bigr)^3 \ , \\[1ex]
S &=&\dsp \frac{\mucrit n}{3 \pi^2 \nQ \xi(x_0(\bar\si))} \ ,
\ea
\label{eps-approx}
\eeq
where all dependence on the quark matter parameters comes through 
the prefactor $S$, which has units of energy. 
$S$ can be explicitly obtained using \eqn{sigbar-def},
\eqn{R-approx}, and \eqn{Z-strangelet} for the $\xi$ function. 
One could informally think of
$S$ as a ``softness'' parameter of the
strangelet crystal EoS: as $S$ increases, the pressure becomes a more
slowly-rising function of energy density.
We expect that softer equations of state
will yield smaller stars with lower maximum masses.
In Table~\ref{tab:S-values} we give the value of $S$ for a range of
values of the parameters of the underlying quark matter EoS.

At low enough pressures, the electrons become nonrelativistic.
Then $\pcell \approx k_{Fe}^5/(15\pi^2m_e)$, and \eqn{eos-const} simplifies to
an analytic expression for the equation of state,
\beq
\ep_{NR} \approx \frac{4 R^3}{3 Z}
  \Bigl(\frac{125 \pi}{9} m_e^3\Bigr)^{1/5}  n \mucrit \pcell^{3/5}
\label{eos-const-NR}
\eeq
This is a reasonable approximation when $k_{Fe}\lesssim m_e$,
i.e. when $\pcell \lesssim m_e^4/(24\pi^2) \approx 0.0003\,\MeV^4$.
However, as we will see below, at the very lowest pressures the
constant potential approximation becomes inaccurate.

\subsubsection{Coulomb potential approximation}
\label{sec:coulapprox}

We can improve on the constant potential approximation by including the
Coulomb energy of the electrons in the calculation of the pressure.
The equation of state is still given by \eqn{eos-const}, but now
the relationship between $\pcell$ and $k_{Fe}$ is modified
by the addition of a Coulomb energy term
(Ref.~\cite{Salpeter:1961zz},~(5)), yielding
\beq
\pcell = p_{e^-}
 - \frac{\al}{5} \Bigl(\frac{Z^2}{18\pi^7}\Bigr)^{1/3} k_{Fe}^4 \ .
\label{eos-Coul}
\eeq
Unlike the constant potential approximation, the Coulomb potential
approximation gives an energy density that goes to a non-zero value
at zero pressure,
\beq
\ep_{\rm Coul}(0) = \mucrit\, n  \frac{2 Z (\al m_e R)^3}{3\pi^2} \ .
\label{eps0-Coul}
\eeq
Comparing with \eqn{e-bigcell} we see that this corresponds to
the energy of cells with size of order $1/(\al m_e) \sim 10^{-10}\,\m$.
This is the energy density of a lattice of zero-pressure atomic matter 
with strangelets in place of nuclei, which is a reasonable guess
for the low-pressure configuration of strangelets. We will therefore
use the Coulomb approximation as the low-pressure extension of
our equation of state. 
As we will see, this leads to a ``planet'' branch
in the mass-radius relation for configurations of strange matter.

\subsubsection{Radius and charge of strangelet at low pressure}
\label{sec:RZ}

The low-pressure approximation expressions given above depend on the
size $R$ and charge $Z$ of the strangelet at the center of a 
large cell. This is approximately an isolated strangelet,
whose radius can be calculated by
minimizing the isolated strangelet free energy given in eqn (25) of 
Ref.~\cite{Alford:2006bx},
\beq
\overline{\De g}(x) =  -\frac{3}{2}\frac{x - \tanh x}{x^3} 
  + \frac{3\bar\si}{x}\ ,
\label{Delta_g_dimless}
\eeq
where $x$ is the radius of the strangelet  in units of $\lambdaD$, 
and 
\beq
\bar\si = \frac{\si}{4\pi\alpha n_Q^2 \lambdaD^3} \ .
\label{sigbar-def}
\eeq
So the strangelet radius $R$ as a function of the parameters of the quark matter
equation of state is
\beq
R = x_0 \lambdaD,\quad \mbox{where}\quad
\frac{d \overline{\De g}}{dx}(x_0) = 0 \ .
\label{R-strangelet}
\eeq
We are interested in values of $\bar\si$ up to 0.13, since for higher
surface tension the strangelet crystal is 
no longer stable \cite{Alford:2006bx}.
An approximate expression for the solution to \eqn{R-strangelet},
accurate to about 0.2\% for  $\bar\si\lesssim 0.13$, is
\beq
x_0^{\rm approx} = \biggl(\frac{15\bar\si}{2}\biggr)^{1/3}
 + \frac{2.174 \,\bar\si}{1-3.982\,\bar\si} \ ,
\label{R-approx}
\eeq
where the first term is the leading-order analytic expression
for $x_0$ in the limit of small $\bar\si$.

The charge $Z$ of the central strangelet is given by eqn.~(17) of 
Ref.~\cite{Alford:2006bx}, which can be written
\beq
\ba{rcl}
Z &\approx&\dsp \frac{4}{3}\pi R^3 \nQ \,\xi(R/\lambdaD) \ , \\[2ex]
\xi(x) &\equiv&\dsp \frac{3}{x^3}(x-\tanh x) \ , 
\ea
\label{Z-strangelet}
\eeq
where $\xi$ is a correction for the effects of screening inside the 
quark matter; it is an even function with $\xi(0)=1$.

\section{Numerical results}
\label{sec:results}

\subsection{Range of parameters studied}
\label{sec:params}

Our assumption that the strange matter hypothesis is valid
requires that $\mucrit$ must be less than
the quark chemical potential of nuclear matter, about 310~MeV,  so we
fix $\mucrit=300~\MeV$. The value of $\mu$ inside our strange matter lumps
will always be within a few MeV of $\mucrit$, because if the surface tension
is small enough to favor the strangelet crystal
it will not cause significant compression.

We will perform calculations for $\la_D=4.82~\fm$ and $\la_D= 6.82~\fm$,
corresponding to $\chiQ \approx 0.2 \mucrit^2$ (appropriate for
unpaired quark matter \eqn{unpaired}) and $\chiQ\approx 0.1 \mucrit^2$
(appropriate for 2SC quark matter \cite{Alford:2006bx}). 

Typical values of $\nQ$ will be around  $0.05\mucrit m_s^2$
\eqn{unpaired}, and a reasonable range would correspond to varying
$m_s$ over its physically plausible range, from about 100 to 300 MeV.
(To have strange matter in the star, $m_s$ must be less than $\mucrit$.)
In this paper we use $\nQ=0.0445$, $0.0791$, and
$0.124~\fm^{-3}$,
which would correspond to $m_s=150$, 200, and 250~MeV in
\eqn{unpaired}.

There is
another widely-discussed phase of quark matter, the color-flavor locked 
(CFL) phase, but it is a degenerate case where $\nQ=\chiQ=0$. CFL strangelets
have a surface charge, but it does not arise from the mechanism studied
here, Debye screening, and has a different dependence on the size
of the strangelet \cite{Madsen:2001fu}. We hope to study CFL strangelet matter
in a separate work.

% Nuclear matter:
%  n_Q = 0.5 n_B = 0.5*0.18 nucleon/fm^3 = 0.09 fm^-3
%  chi_Q = mu_B^2/pi^2 = 2.3 fm^-2
% mue =  nQ/chiQ   should be much bigger!

\begin{figure}[htb]
\includegraphics[width=\hsize]{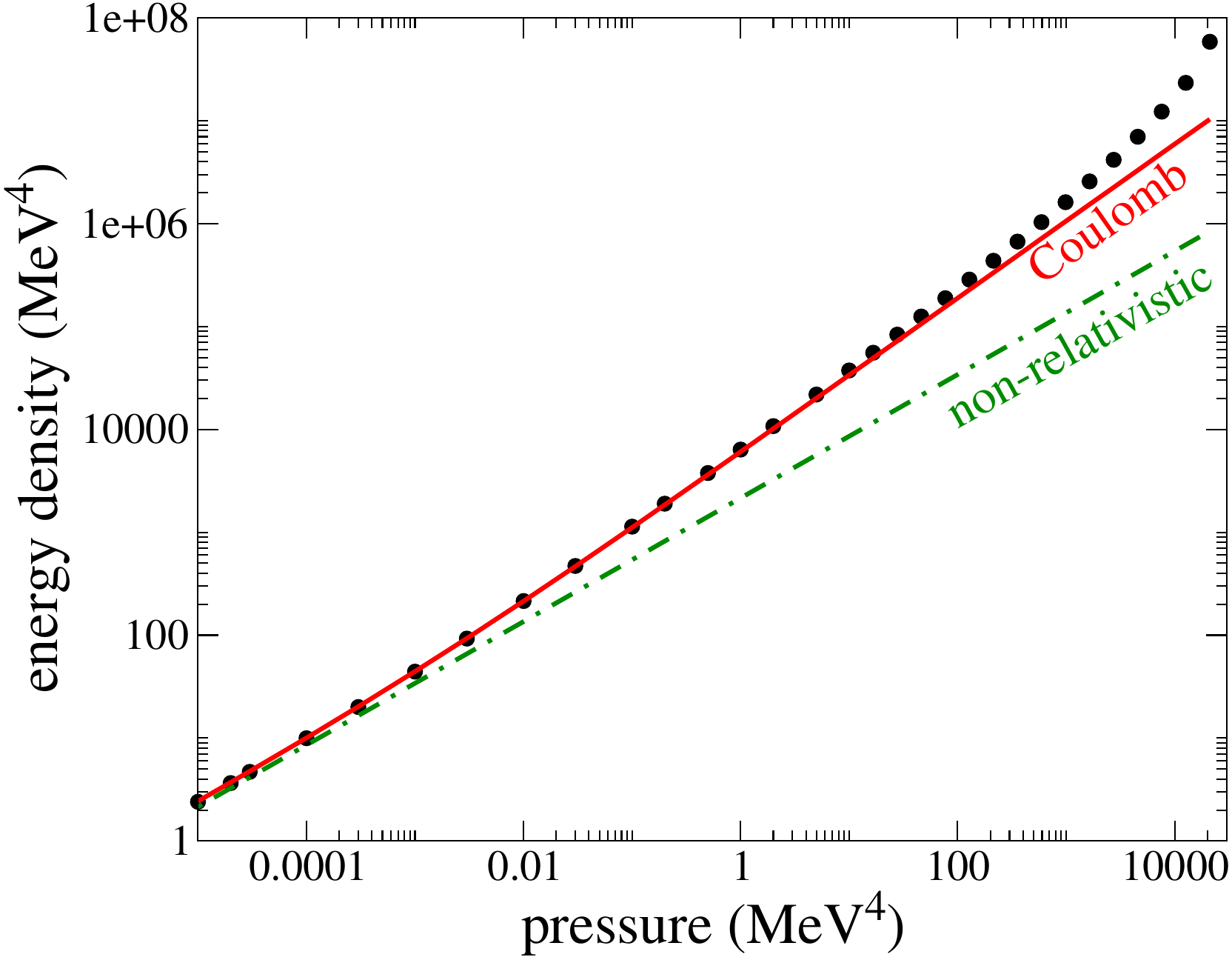}
\caption{
Equation of state of the mixed phase (strangelet crystal) for strange matter
with $\mucrit=300\,\MeV$,
$\lambdaD=6.82\,\fm$, $\protect\nQ=0.0791\,\fm^{-3}$,
$\sigma=1.0\,\MeV\fm^{-2}$.
The dots were obtained numerically following the procedure of
Sec.~\ref{sec:exact}. The solid line is the Coulomb-potential
approximation (Sec.~\ref{sec:coulapprox}). 
The dashed line is the non-relativistic electron
(ultra-low pressure) limit \eqn{eos-const-NR}.
Above $p\approx 20000\,\MeV^4$, uniform quark matter becomes favored
over the mixed phase.
}
\label{fig:eos}
\end{figure}

\begin{figure}[htb]
\includegraphics[width=\hsize]{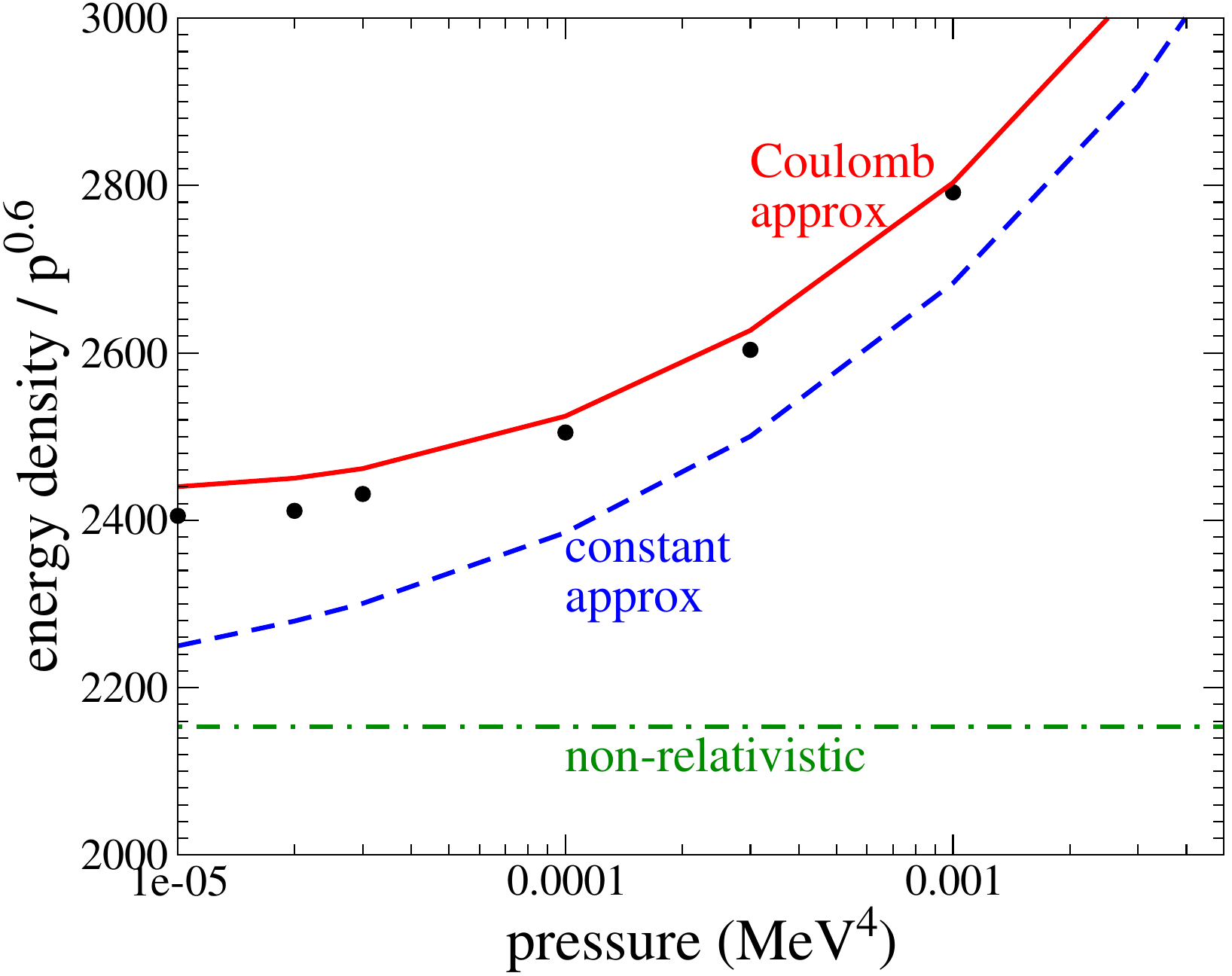}
\caption{
Equation of state of the mixed phase for the same parameters as 
in Fig.~\ref{fig:eos}, zoomed in on the low pressure region, and with
the energy density divided by $p^{0.6}$. 
The dots were obtained numerically following 
the procedure of Sec.~\ref{sec:exact}. The Coulomb-potential
approximation (Sec.~\ref{sec:coulapprox}) is the most accurate,
followed by the constant-potential approximation (Sec.~\ref{sec:constapprox}),
and then the  non-relativistic electron
approximation \eqn{eos-const-NR}.
}
\label{fig:eos_ratio}
\end{figure}

\begin{figure}[tbh]
\includegraphics[width=\hsize]{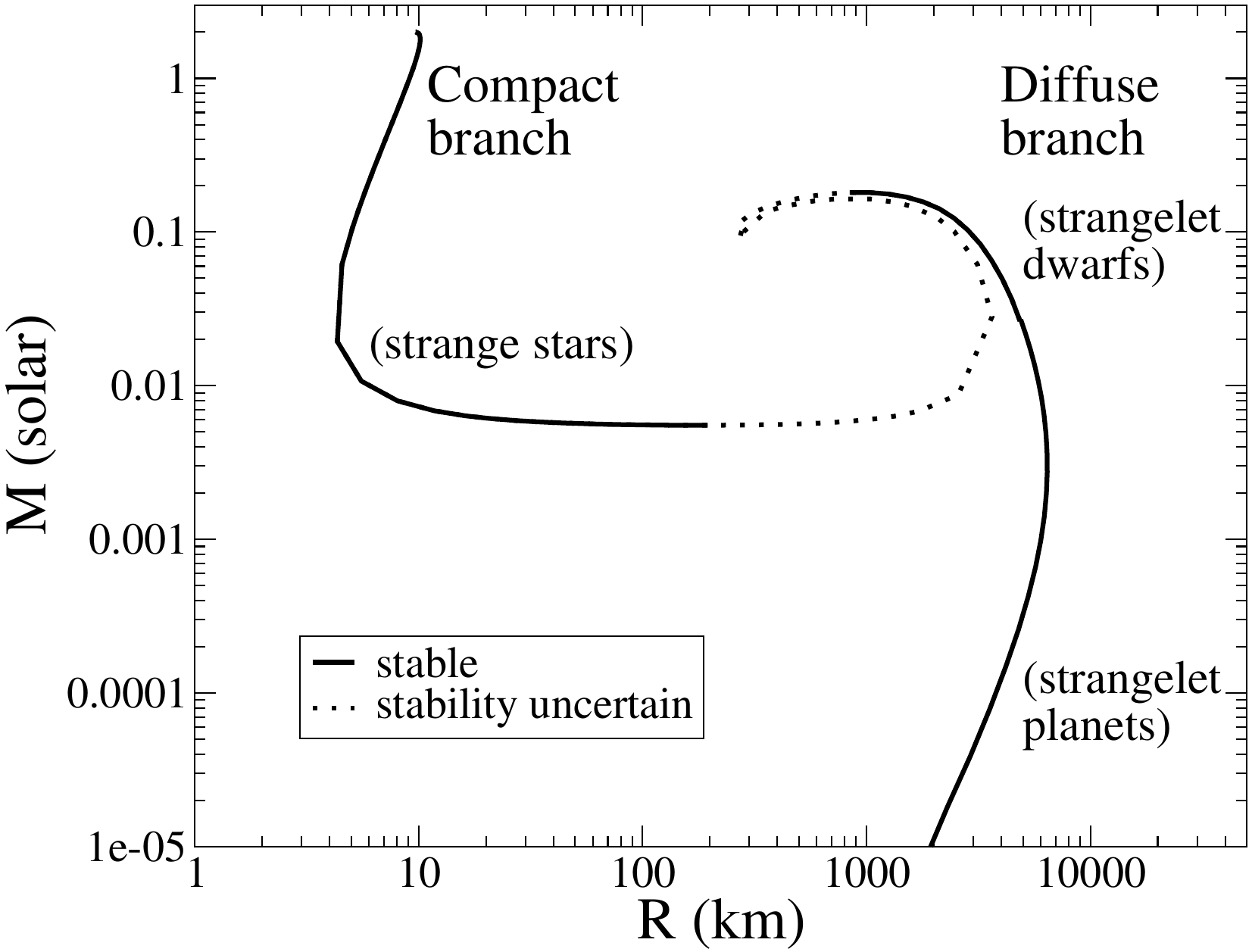}
\caption{
The full mass-radius curve for stars made of quark matter with the equation
of state plotted in Fig.~\ref{fig:eos}, using the Coulomb approximation 
\eqn{eos-Coul} to extrapolate to lower pressures.
% $\mucrit=300\,\MeV$, $\lambdaD=6.82\,\fm$,  $\protect\nQ=0.0791\,\fm^{-3}$,
% $\sigma=1.0\,\MeV\fm^{-2}$. 
The compact branch contains strange stars
with a strangelet crystal crust. The diffuse branch contains stars
consisting entirely of strangelet crystal matter. Solid lines
represent configurations that are stable; stability of the other branches
is discussed in the text. 
}
\label{fig:MR-full}
\end{figure}

\begin{figure}
\includegraphics[width=\hsize]{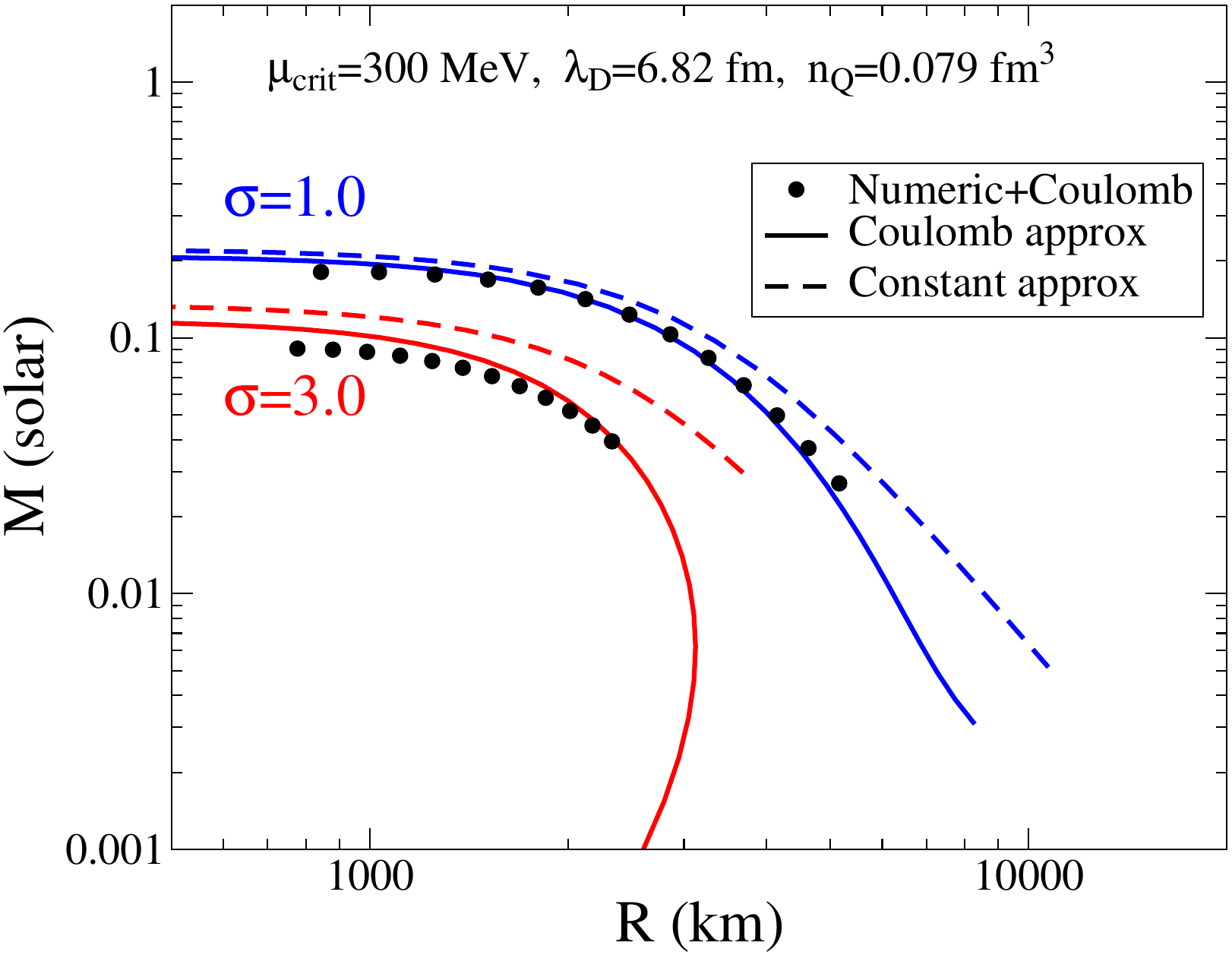}
\caption{
Mass-radius relation for strangelet dwarfs made of strangelet crystal
matter, comparing different approximations to the equation of state.
Upper (blue) curves are for the same parameters as in Figs.~\ref{fig:eos}
and \ref{fig:eos_ratio}. Lower (red) curves are for
a larger surface tension, $\si=3\,\MeV\fm^{-2}$. 
The dots were obtained using the full numerical
equation of state (Sec.~\ref{sec:exact}). The solid lines
use the Coulomb-potential approximation (Sec.~\ref{sec:coulapprox}), and the
dashed lines use the constant-potential approximation \eqn{eos-const}.
}
\label{fig:MR-varysigma}
\end{figure}

\begin{figure}
\includegraphics[width=\hsize]{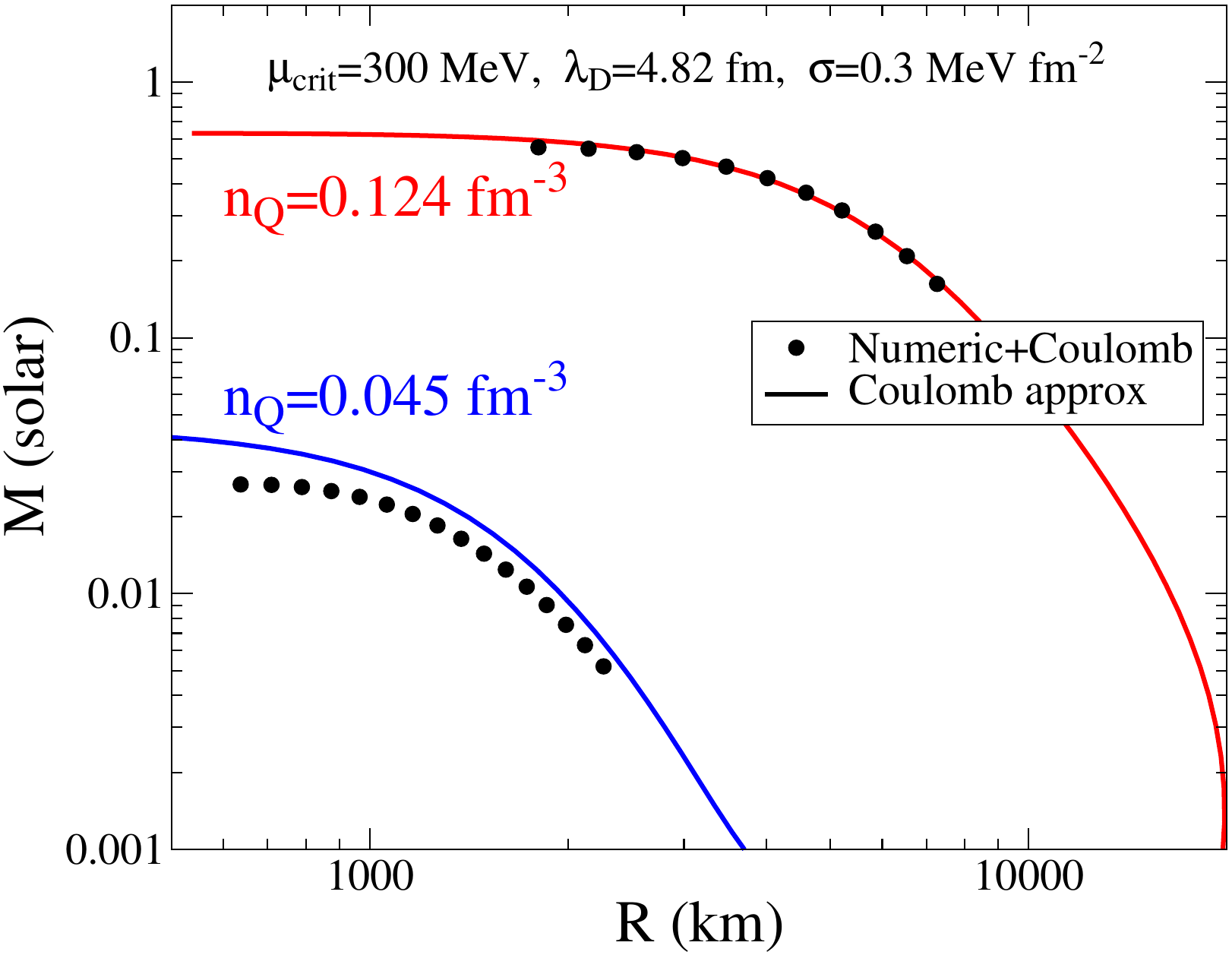}
\caption{
Mass-radius relation for strangelet dwarfs made of strangelet crystal
matter, comparing different approximations to the equation of state.
%Upper (blue) curves are for the same parameters as in Figs.~\ref{fig:eos}
%and \ref{fig:eos_ratio}. Lower (red) curves are for
%a larger surface tension, $\si=3\,\MeV\fm^{-2}$. 
%The dots were obtained using the full numerical
%equation of state (Sec.~\ref{sec:exact}). The solid lines
%use the Coulomb-potential approximation (Sec.~\ref{sec:coulapprox}), and the
%dashed lines use the constant-potential approximation \eqn{eos-const}.
}
\label{fig:MR-varynQ}
\end{figure}

\begin{table*}
\setlength{\tabcolsep}{1em}
\begin{tabular}{cccccccc}
$\lambda_D$ & $n_Q$ & $\sicrit$  
& \multicolumn{4}{c}{Softness prefactor $S$ (MeV) at}  \\
   (fm)     & (fm${}^{-3}$) & (MeV fm${}^{-2}$) & $\si\!=\!0.3$ &  $\si\!=\!1.0$ 
  & $\si\!=\!3.0$ &  $\si\!=\!10.0$ &  \\[1ex]
\hline
4.82  & 0.0445 & 0.533 & 345  & -- & -- & --  \\ 
4.82  & 0.0791 & 1.69  & 158 & 202 & -- & --  \\
4.82  & 0.124 & 4.12   & 94  & 104 & 140 & --  \\[1ex]
6.82  & 0.0445 & 1.51 & 280  & 367  &  -- & -- \\
6.82  & 0.0791 & 4.8  & 146  & 161 & 206 & -- \\
6.82  & 0.124 & 11.6  & 90  & 95 & 105 & 155 \\[1ex]
\hline
\end{tabular}
\caption{
Softness prefactor \eqn{eps-approx} of the strangelet crystal for various
quark matter equation of state.
The first two columns, $\la_D$ and $\protect\nQ$,
specify the quark matter equation of state \eqn{generic_EoS}
(via \eqn{Debye_length}). The third column gives the maximum surface tension
for which a strangelet crystal will occur \eqn{sicrit_result}.
The last four columns give the softness prefactor $S$ for different values
of the surface tension  $\si$ (given in MeV\,fm${}^{-2}$)
of the interface between quark matter and vacuum.
} 
\label{tab:S-values}
\end{table*}

% \clearpage

\subsection{Testing approximations to the equation of state}

In Fig.~\ref{fig:eos} we show the equation of state for the
strangelet crystal, for critical quark chemical potential $\mucrit=300\,\MeV$,
quark matter screening distance $\lambdaD=6.82\,\fm$,
quark charge density parameter $\nQ=0.0791\,\fm^{-3}$,
and quark matter surface tension $\sigma=1.0\,\MeV\fm^{-2}$.
The dots were obtained numerically following the procedure of
Sec.~\ref{sec:exact}. The solid line is the Coulomb-potential
approximation (Sec.~\ref{sec:coulapprox}). On this
plot the constant potential
approximation (Sec.~\ref{sec:constapprox}) line would be indistinguishable 
from the Coulomb-potential line, so we do not show it.
The dot-dashed line is the non-relativistic electron
(ultra-low pressure) limit \eqn{eos-const-NR} of the constant potential
approximation.
Above $p\approx 20000\,\MeV^4$, uniform quark matter becomes favored
over the mixed phase.
On this very expanded logarithmic scale, the Coulomb approximation
appears reasonably
accurate up to pressures of order 1\,MeV.

To achieve more discrimination between the different approximations, we 
show in Fig.~\ref{fig:eos_ratio} a magnified version of the low-pressure
end of the plot in Fig.~\ref{fig:eos}, where we have divided out the
non-relativistic scaling of the energy density, $\varepsilon \sim p^{3/5}$.
We can see that, down to the lowest pressures for which we can
perform the numerical Wigner-Seitz calculation of the equation of state,
the Coulomb approximation gives the most accurate semi-analytic
approximation, although the constant potential approximation is
accurate to within about 10\%.

We then have to decide which approximation to use for lower
pressures, where numerical calculations are not available.
In the low-pressure limit, the Coulomb approximation to $\ep(p)$ tends to a
fixed value, while the constant and nonrelativistic approximations to $\ep(p)$
tend to zero as $p^{3/5}$. So in Fig.~\ref{fig:eos_ratio} the Coulomb
approximation will diverge at $p\ll 10^{-5}\,\MeV^4$, while the constant and
nonrelativistic approximations will tend to the same constant value. 
As discussed in Sec.~\ref{sec:coulapprox}, it seems reasonable to expect that at
the lowest pressures there will be a crystal of ``strange atoms'', each
consisting of electrons bound to a strangelet, and the Coulomb approximation
gives a reasonable estimate of the energy density of such matter, 
so at low pressure we will
use the Coulomb approximation.

\subsection{Mass-radius relation of strange stars}

% The mass-radius relation for strangelet dwarfs is very sensitive to
% the low-pressure equation of state. 

In Fig.~\ref{fig:MR-full} we show
the full mass-radius curve for stars made of quark matter with the equation
of state plotted in Fig.~\ref{fig:eos}.
% $\mucrit=300\,\MeV$, $\lambdaD=6.82\,\fm$,  $\protect\nQ=0.0791\,\fm^{-3}$,
% $\sigma=1.0\,\MeV\fm^{-2}$. 
The compact branch contains strange stars
with a strangelet crystal crust. The diffuse branch contains stars
consisting entirely of strangelet crystal matter. It includes
two segments: the lighter one is planets of dilute
strange matter whose the mass increases with radius.
This joins to the strangelet dwarf
branch where the mass decreases with radius as the strangelet crystal
is compressed
by the pressure due to gravity.  We use the numerically
calculated equation of state (Sec.~\protect\ref{wigner-seitz}) except that
at very low pressure (the planetary branch) the Wigner-Seitz cells
become so large that our numerical methods break down, so as
discussed in Sec.~\ref{sec:coulapprox} we
use the Coulomb approximation \eqn{eos-Coul} to extrapolate
down to zero pressure.
% which we have already seen is reasonably accurate (Fig.~\ref{fig:eos_ratio}).

Fig.~\ref{fig:MR-full} shows the whole $M(R)$ curve, not all of which
corresponds to stable configurations. 
The usual stability criterion for stars \cite{Bardeen:1966} is that
one radial mode becomes either stable or unstable at each extremum in the
$M(R)$ function. A stable mode
becomes unstable at each extremum where
the curve bends counterclockwise as the central density increases;
a stable mode becomes unstable at each extremum where the curve bends 
clockwise as the central density increases. However, 
Glendenning~et.al.~\cite{Glendenning:1994sp}
report that at some extrema there is no change in stability: the squared
frequency of one of the fundamental radial modes may touch zero, but
not change sign.
We defer a detailed study of the stability of radial modes of 
strange stars to future work, and in Fig.~\ref{fig:MR-full} we
show as ``stable'' (solid curves) the parts of the $M(R)$ curve that both
Ref.~\cite{Bardeen:1966} and Ref.~\cite{Glendenning:1994sp} agree are stable.
We note that Ref.~\cite{Glendenning:1994sp} is a study of
stars that have a core of uniform strange matter surrounded
by a crust of nuclear matter: these are 
similar to the configurations along the dashed part
of the mass-radius curve in Fig.~\ref{fig:MR-full}, where we have a core
of uniform strange matter surrounded by a crust of strangelets, with a
density discontinuity at the boundary. If Ref.~\cite{Glendenning:1994sp}'s
stability argument is correct and applicable to our stars, 
then some of these configurations may also be stable.
In the remainder of this paper we will focus on the strangelet dwarf
branch, which consists of a simple crystal of stranglets with no
uniform core, so there is no controversy about the appropriate
stability criterion.

\subsection{Mass-radius relation of strangelet dwarfs}

To investigate the sensitivity of the masses and radii of strangelet dwarfs to
the parameters of the quark matter equation of state, we show in 
Fig.~\ref{fig:MR-varysigma} and ~\ref{fig:MR-varynQ} 
the strangelet dwarf part of the 
mass-radius curve, excluding the compact and planetary branches, for
various values of the quark matter parameters.

In Fig.~\ref{fig:MR-varysigma} we explore the effects of varying
the surface tension, and we compare the different approximations to
the equation of state.
The upper curves are for the same equation of state
as was shown in Fig.~\ref{fig:eos}
and \ref{fig:eos_ratio}; the lower curves use a larger surface tension,
$\sigma=3\,\MeV\fm^{-2}$. In both cases the solid curves are obtained from the
Coulomb-potential approximation to the equation of state,
and the dashed lines are obtained from the constant-potential
approximation. The dots use the equation of state that
is obtained numerically following the procedure of
Sec.~\ref{sec:exact}, except that at very low pressures, where the
numerical calculation becomes too difficult, the Coulomb approximation
is used.

We see that, as one might have expected from Fig.~\ref{fig:eos},
using the Coulomb approximation over the entire pressure range of the
mixed phase yields reasonably accurate results.
However, as noted in Sec.~\ref{sec:constapprox},
the constant potential approximation is still useful
for gaining an understanding of how the $M(R)$ curve for strangelet dwarfs 
depends on the parameters of the EoS, because in the range of
pressures that is important for strangelet dwarfs it gives
a good indication of the $M(R)$ curve. 
(At ultra-low pressures,
relevant for the strange planet branch, this is no longer the case: one has
to use the Coulomb approximation instead.)
As discussed in Sec.~\ref{sec:constapprox}, the constant potential
approximation to the EoS can be written in terms of a ``softness prefactor''
$S$ \eqn{eps-approx}. To understand how the $M(R)$ curve in 
Fig.~\ref{fig:MR-varysigma} changes with $\si$, note that 
$x_0(\bar\si)$ is a monotonically increasing
function and $\xi(x_0)$ is a monotonically decreasing function,
so as the surface tension $\si$ increases at fixed values
of the other parameters, the softness prefactor $S$ of the
strangelet crystal EoS increases 
(one can see this in Table~\ref{tab:S-values}). Since the
EoS is becoming softer, the $M(R)$ curve moves down
and to the left, giving smaller stars with a lower maximum mass.

In Fig.~\ref{fig:MR-varynQ} we explore the effects of varying
the charge density parameter $\nQ$ in \eqn{generic_EoS}
while keeping the other parameters constant. 
As in Fig.~\ref{fig:MR-varysigma}, solid lines are for the Coulomb approximation
to the equation of state, dots are for the numerically calculated 
equation of state using the Coulomb approximation to extrapolate to the
lowest pressures. We see that increasing $\nQ$  yields 
heavier, larger strangelet dwarf stars. Again, this can be understood
in terms of the constant potential approximation and its softness
prefactor $S$ \eqn{eps-approx}. As $\nQ$ increases, it causes
$S$ to decrease through two effects. Firstly via
the explicit factor of $\nQ$ in the denominator of \eqn{eps-approx},
and secondly via the relationship \eqn{sigbar-def} between 
$\si$ and $\bar\si$. The sensitivity of $S$ to changes in $\nQ$ can
be seen in Table~\ref{tab:S-values}: for the two values of $\nQ$
used in  Fig.~\ref{fig:MR-varynQ} the values of $S$ are near the
extremes of its range in the parameter set we studied: $S\approx 345$
and $S\approx 94$ for $\nQ=0.0445$ and $\nQ=0.124$ respectively.
Consequently, the $M(R)$ curve for $\nQ=0.0445$ is characteristic of
a soft equation of state, with low radius at a given mass and a low 
maximum mass, whereas the $M(R)$ curve for $\nQ=0.124$ is characteristic of
a hard equation of state, with large radius at a given mass and a high
maximum mass.

\section{Discussion}
\label{sec:discussion} 

We have shown that, if the strange matter hypothesis is correct
and the surface tension of the interface between strange matter 
and the vacuum is less
than a critical value \eqn{sicrit_result}, there is at least one
additional stable branch in the mass-radius relation for strange stars,
corresponding to large diffuse objects that we call ``strangelet dwarfs'',
consisting of a crystal of strangelets in a sea of electrons.
This is easily understood, since if $\si<\sicrit$ then uniform strange matter
is unstable at zero pressure, and undergoes
charge separation to a crystal of positively-charged
strangelets surrounded by electrons,
just as normal matter at zero pressure is a mixed phase 
consisting of droplets of
nuclear matter surrounded by electrons. Strangelet dwarfs are then
the strange matter equivalent of white dwarfs.

We emphasize that in this low-surface-tension scenario, strange matter
is {\em not} self bound. Like nuclear matter, it is only bound by
gravitational forces. Every strange star will have a strangelet crystal
crust, and strangelet dwarfs are those strange stars that are ``all crust''.

The natural production mechanism by which strangelet dwarfs might
be produced is a collision between a strange star and another compact
object. In such collisions, up to $0.03\,M_\odot$
may be ejected \cite{Bauswein:2008gx}, which is in the mass range we are
predicting for strangelet dwarfs. There are two ways a collision could
produce strangelet dwarfs. Firstly, part of the crust of the strange star
might be ejected to become a isolated object, which would be a
strangelet dwarf. Secondly, if a sufficiently light piece of 
the uniform quark matter core were ejected in the collision, 
it would be unable to
exist on the compact branch, and would evaporate into
a configuration on the diffuse branch. For example, 
for the equation of state studied in Fig.~\ref{fig:MR-full}, the
lightest compact configuration of strange matter is $0.0055\,M_\odot$.
A lighter piece of strange matter could only exist on the diffuse branch,
and would spontaneously evaporate to become a strangelet dwarf.
Strangelet dwarfs produced by these mechanisms could then bind 
gravitationally, to form heavier strangelet dwarfs.

It should be noted that our proposed mechanism for the production
of strangelet dwarfs is also a mechanism for creating a diffuse cosmic
flux of strangelets (``strangelet pollution''), which might be expected to 
convert all neutron stars
to strange stars \cite{Friedman:1990qz}. Although observations of
glitches and magnetar oscillations \cite{Watts:2006hk} seem consistent
with some compact stars having nuclear matter crusts, there remains some
uncertainty. Crystalline phases of quark matter could allow strange stars
to glitch \cite{Mannarelli:2007bs}, and in our low-surface-tension scenario
strange stars have crusts that could be hundreds of meters thick 
\cite{Alford:2008ge}. A cosmic flux of strangelets may seem unlikely but until
it is ruled out experimentally (as may happen soon from the AMS experiment
\cite{Sandweiss:2004bu}) it remains useful to analyze the full observational
consequences of the strange matter hypothesis.

Our analysis assumes that at any
given pressure the strangelet crystal consists of the most energetically
favorable strangelet configuration (in terms of strangelet size and charge
and cell size). However, other configurations will in general be metastable
with long lifetimes. If one compresses a piece of strangelet crystal then
the charge of the strangelets can readily change via absorption or
emission of electrons, but it is very difficult for the
quark matter to rearrange itself in to strangelets of the 
now-energetically-favored size: it is more likely that
the strangelets will stay the same size and the radial density profile
of the electrons will change. The sizes of the strangelets will
be determined more by the history of the object than by the pressure.
Taking this point further, it is quite possible to have a 
crystal consisting of a mixture of strangelets and ordinary nuclei,
held apart by their electrostatic repulsion but also bound together
in to a crystal by the degenerate electron gas that neutralizes them,
forming a hybrid strangelet/white dwarf star.

Detection of strangelet dwarfs requires an observation method that can find
non-luminous objects with typical masses of $10^{-5}$ to $10^{-1}\,M_\odot$ 
and radii in the range $500$ to $5000$\,km. An example is
gravitational microlensing surveys, such as those conducted by
the Microlensing Observations in Astrophysics (MOA) and the Optical
Gravitational Lensing Experiments (OGLE) groups, which look for
lensing events in the galactic bulge, and are capable of detecting
Jupiter-mass objects.  It is intriguing that such surveys now
report the existence of an abundant population of unbound distant planetary
masses, suggesting that such objects may be twice as common as main
sequence stars \cite{Sumi:2011}. Although models of planet formation 
indicate that mechanisms exist for unbinding planets through disk instabilities
and planet interactions \cite{Veras:2009}, we suggest that a possible
alternative is formation of strange dwarfs from matter ejected in strange 
star mergers. One would expect that sometimes a  strangelet dwarf produced
in a merger might be unable to escape the gravitational field of
the remaining compact object, and this would explain the presence of
dense planet-mass objects in the vicinity of compact stars. An example is
the millisecond pulsar PSR J1719-1438, which has a Jupiter-mass
companion whose inferred central density 
($\rho > 23\,{\rm g}\,{\rm cm}^{-3}$) is
far in excess of what is expected in a planet \cite{Bailes:2011}.
We expect that in the near future further light will be cast on this question,
as microlensing surveys help us better
understand the distribution of planetary mass compact objects and
as strategies are devised to provide information about both mass and radius.

\section*{Acknowledgments}

We thank E. Agol and F. Weber for useful discussions. 
This research was
supported in part by the 
Offices of Nuclear Physics and High Energy Physics of the
Office of Science of the 
U.S.~Department of Energy under contracts
\#DE-FG02-91ER40628,  % Wash U high energy theory
\#DE-FG02-05ER41375, % Mark's nuclear theory
and by the DoE Topical Collaboration 
``Neutrinos and Nucleosynthesis in Hot and Dense Matter'', 
contract \#DE-SC0004955.
%and by the Undergraduate Research
%Office and McDonnell Center for Space
%Sciences of Washington University in St. Louis.

% Override the revtex href command in order that the JHEP bib style
% will work properly:
\renewcommand{\href}[2]{#2}

% macros used by ADS Database BiBTeX entries:
% see http://adsabs.harvard.edu/abs_doc/aas_macros.sty
\newcommand{\apjl}{Astrophys. J. Lett.\ }
\newcommand{\mnras}{Mon. Not. R. Astron. Soc.\ }
\newcommand{\aap}{Astron. Astrophys.\ }

\bibliographystyle{JHEP_MGA}
\bibliography{dwarf} 

\end{document}